# Coupling between hyperbolic phonon polaritons excited in two ultrathin hBN sheets

Xiaohu Wu[1,*], Chengshuai Su[2], Kezhang Shi[3], Feng Wu[4], and Ceji Fu[2]

[1]Shandong Institute of Advanced Technology, Jinan 250100, China

[2]LTCS and Department of Mechanics and Engineering Science, College of Engineering, Peking University, Beijing 100871, China

[3]Centre for Optical and Electromagnetic Research, State Key Laboratory of Modern Optical Instrumentation, Zhejiang University, Hangzhou 310058, China

[4]School of Optoelectronic Engineering, Guangdong Polytechnic Normal University, Guangzhou 510665, China

Corresponding author: xiaohu.wu@iat.cn

In this work, the hyperbolic phonon polaritons (HPPs) in ultrathin hBN sheets are numerically studied. The dispersion relation and distribution of electric field are calculated to confirm the excitation of HPPs. Besides, the coupling effect between HPPs of two ultrathin hBN sheets are investigated. When the distance between two hBN sheets are smaller than the propagation length of the HPPs in the air, the HPPs can be strongly coupled. Therefore, the photon tunneling probability can be greatly enhanced. The split of the HPPs is similar to that of the surface waves, and such phenomenon is well explained in this work. We believe that this work will deepen our understanding of the HPPs in ultrathin hyperbolic materials. In addition, the knowledge about the HPPs will help us understand the near-field radiative heat transfer between hyperbolic materials.

probability, hBN.

## 1. Introduction

Over the past two decades, hyperbolic materials have attracted researchers' great interest due to their applications in optical absorbers [1, 2], optical mirrors [3], and super-resolution imaging [4, 5]. Recently, researchers discovered a kind of special phonon polaritons called hyperbolic phonon polaritons (HPPs) in hyperbolic materials [6-8]. Hexagonal boron nitride (hBN) is a representative hyperbolic material, in which HPPs possess extremely high confinement and low loss [9-14]. The natural hyperbolicity makes hBN good candidate for near-field optical imaging and focusing [13]. However, the HPPs in ultrathin hBN sheets have not been fully considered and remain elusive [14].

Particularly, previous works have shown that the coupled HPPs between two hyperbolic plates can enhance the photon tunneling probability, and it can be used in near-field radiative heat transfer to improve the energy transfer efficiency and realize super-Planckian thermal radiation [15-27]. However, when the thickness of hBN is very small, the coupling of HPPs in hBN is similar to that of surface waves, and thus the HPPs was considered as surface phonon polaritons [19]. It is noted that the dispersion relations for hyperbolic volume phonon polaritons and hyperbolic surface phonon polaritons are possible to be identical in ultrathin hyperbolic sheets, thus the dispersion relations cannot be solely treated as evidence to confirm the types of phonon polaritons excited in the media [24,28]. Therefore, it is urgent to clarify the HPPs in ultrathin hBN sheets and it can further help us understand the coupling of HPPs between two hBN sheets.

In this work, taking hBN for an example, the HPPs in ultrathin hBN sheets are investigated. Moreover, the coupling effect between HPPs in ultrathin hBN sheets are studied. The dispersion relations and distribution of electric field are calculated to identify the types of phonon polaritons excited in the ultrathin hBN sheets. The results obtained in this work will deepen our understanding about the HPPs in ultrathin hyperbolic materials.

## 2. Results and discussion

hBN is a kind of natural hyperbolic materials whose permittivity tensor can be expressed as $\mathrm{diag}(\varepsilon_\perp,\varepsilon_\perp,\varepsilon_\parallel)$ when its optical axis is along $z$-axis. The explicit expression of $\varepsilon_\perp$ and $\varepsilon_\parallel$ can be found in Ref. [9]. The hBN has two hyperbolic bands, the one between $1.47\times10^{14}$ rad/s and $1.56\times10^{14}$ rad/s is called the type I hyperbolic band in which $\mathrm{Re}(\varepsilon_\perp)>0$ and $\mathrm{Re}(\varepsilon_\parallel)<0$, and the other between $2.58\times10^{14}$ rad/s and $3.03\times10^{14}$ rad/s is called the type II hyperbolic band in which $\mathrm{Re}(\varepsilon_\perp)<0$ and $\mathrm{Re}(\varepsilon_\parallel)>0$. The plot of the permittivity can be found in lots of published literatures [22,25]. The structure considered in this work is shown in Fig. 1, where the thickness of each hBN sheet is $h$ and the distance between them is $d$.

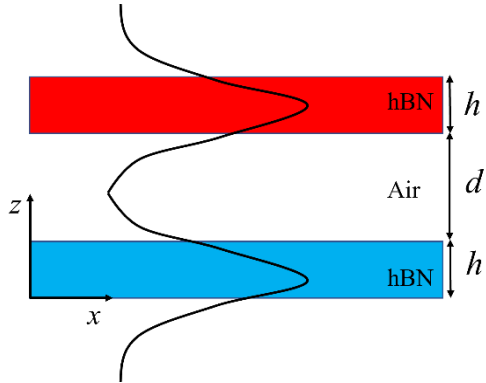

Fig. 1 Schematic of the coupling between two hBN sheets. The distance between them is *d* and their thicknesses are both *h*.

The HPPs in each hBN sheet can be found by investigating the imaginary part of the $r_p$ (the reflection coefficient of the hBN sheet for an incident transverse magnetic (TM) wave), as shown in Fig. 2(a). The thickness of each hBN sheet is 5 nm, and two orders of HPPs can be seen. The 0-roder HPP is much strong, while the 1-order HPP is weak. The dispersion of HPPs in thin hBN can be derived using the Fabry-Perot quantization condition, and the analytical approximate relation is [9,28]

$$\beta = \frac{\rho}{k_0 h}\left[2\arctan\left(\frac{\rho}{\varepsilon_\parallel}\right) + \pi l\right], \quad l = 0,1,2,... \tag{1}$$

where $\beta$ is the dimensionless wavevector component parallel to the surface, $\rho$ is defined as $i\sqrt{\varepsilon_\parallel/\varepsilon_\perp}$, $k_0$ is the wavevector in vacuum. As shown in Fig. 2(a), it is clear the dispersion relations agree very well with the simulation. Therefore, the dispersion relation can partially confirm the excitation of HPPs.

Previous works have shown that the coupling HPPs between two hBN films can enhance the near-field radiative heat transfer. The photon tunneling probability can be used to describe the coupling and it can be expressed as [25]

$$\xi(\omega,\beta) = 4\frac{\left[\text{Im}(r_p)\right]^2 e^{-2|k_z|d}}{\left|1 - r_p^2 e^{-2ik_z d}\right|^2}, \quad \beta > k_0 \tag{2}$$

where $k_z = \sqrt{k_0^2 - \beta^2}$ is the wavevector in the air along the *z*-axis. It is noted that only the contribution of TM waves is included, since the HPPs discussed here can only be excited for TM waves. Besides, the HPPs cannot be excited when $\beta$ is smaller than $k_0$.

When the distance between two hBN sheets is 20 nm, the photon tunneling probability is shown in Fig. 2(b). One can see that the 0-order HPPs will split into two branches, while the 1-order HPPs disappears. The splitting phenomenon is similar to the coupled surface plasmon polaritons between two graphene sheets (see Fig. 4 in Ref. [25]). However, such a splitting does not mean the original modes are always the surface modes.

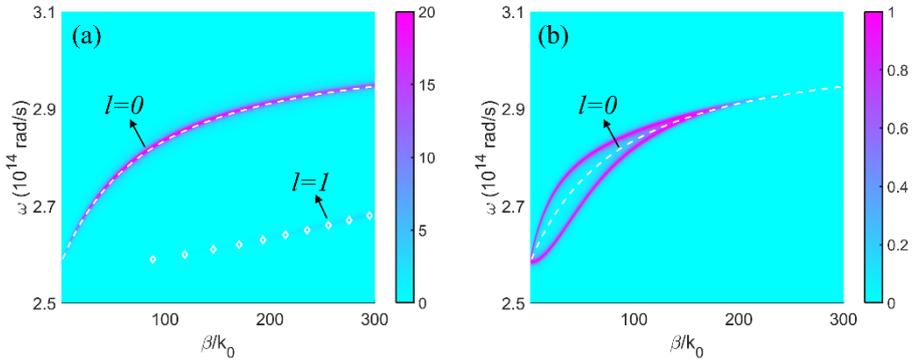

Fig. 2 (a) Imaginary part of $r_p$ as a function of the wavevector and the angular frequency when the thickness of the hBN sheet is 5 nm. (b) Photon tunneling probability as a function of the wavevector and the angular frequency, the thickness of each hBN sheet and the distance between them are 5 nm and 20, respectively. The white dashed line corresponds to the 0-order HPPs, while the white discrete diamond corresponds to the 1-order HPPs.

According to Fig. 2(a), the excitation wavevector is $74k_0$ at angular frequency of $2.8 \times 10^{14}$ rad/s ($\varepsilon_\perp = -5.59 + i0.23$, $\varepsilon_\parallel = 2.80$). To further confirm the excitation of HPPs, instead of surface waves, in the hBN sheet, the distribution of electric field along the $x$-axis is shown in Fig. 3. The electric field is normalized to the electric field at the upper air/hBN interface, and the method is transfer matrix method [29]. The electric field exponentially decays in the air, while it is enhanced in the hBN sheet.

Therefore, the distribution of electric field can ensure that modes excited at the air/hBN interfaces are not surface modes. Both the dispersion relation and distribution of electric field can prove that the hBN sheet with thickness of 5 nm can support 0-order and 1-order HPPs. Besides, the coupling between 0-roder HPPs can enhance the photon tunneling probability when the distance is 20 nm.

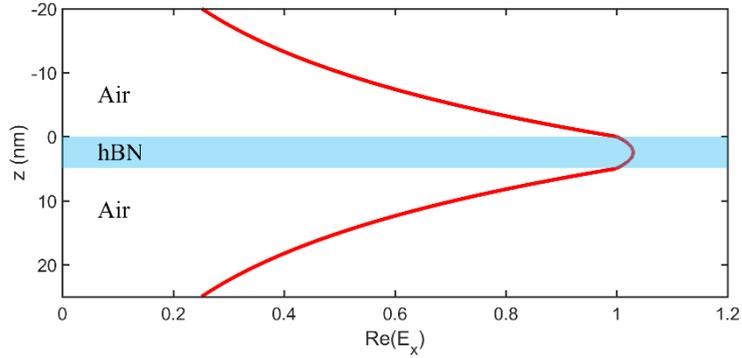

Fig. 3 (a) Normalized amplitude distribution of electric field along the *x*-axis. The angular frequency and wavevector are $2.8 \times 10^{14}$ rad/s and $74 k_0$, respectively.

When the thickness is 50 nm, Im($r_p$) is as shown in Fig. 4(a). One can see that more orders of HPPs can be excited in the hBN film. Besides, the dispersion relations agree very well with the simulation. Fig. 4(b) shows the photon tunneling probability when the distance is 20 nm. Interestingly, the 0-order HPPs are strongly coupled, resulting in the splitting. However, the 1-order HPPs are weakly coupled, thus the splitting is not so obvious. HPPs of higher orders contribute much less to the high-value photon tunneling probability due to the larger wavevector at a same angular frequency.

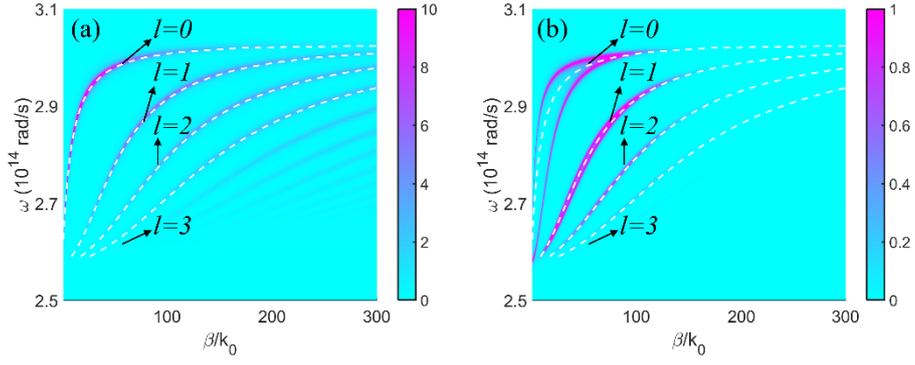

Fig. 4 (a) Imaginary part of $r_p$ as a function of the wavevector and the angular frequency when the thickness of the hBN sheet is 50 nm. (b) Photon tunneling probability as a function of the wavevector and the angular frequency, the thickness of each hBN sheet and the distance between them are 50 nm and 20 nm, respectively. The white dashed lines correspond to different orders of HPPs.

To explain the contribution of HPPs of different orders to the photon tunneling probability, the distribution of electric field along the *x*-axis for different orders of HPP is shown in Fig. 5. The angular frequency is fixed at $2.8 \times 10^{14}$ rad/s, and the corresponding wavevectors for 0-order, 1-order, and 2-order of HPPs are $7.5k_0$, $55k_0$, and $101k_0$, respectively. One can see that the electric field exponentially decays in the air with different ratio. The 0-order HPP decays moderately, while 1-order and 2-order decays dramatically. The propagation length in the air can be defined as $1/|k_z|$. Therefore, the corresponding propagation length in the air for 0-order, 1-order, and 2-order of HPPs are 142.8 nm, 19.5 nm, and 10.6 nm, respectively. The 0-order HPP can propagate larger distance in the air. When putting two hBN sheets at the distance of 20 nm, the coupling between 0-order HPP is much strong, while that between 1-order and 2-order HPPs are weak. The distribution of electric field can well explain the coupling between different orders of HPPs. Besides, the electric field oscillates in the hBN sheet when the order is 2, indicating the excitation of HPPs. The electric field for 0-order HPP in the hBN sheet does not oscillate because the thickness is not thick enough. It is noted that the surface phonon polaritons can be excited at the surface of an hBN sheet of large enough thickness, with wavevector satisfying $k_0 < \beta < \sqrt{\varepsilon_\parallel} k_0$ [29]. We have checked that the electromagnetic waves in the hBN sheets is propagating waves when the wavevector along the $x$-axis is $7.5k_0$.

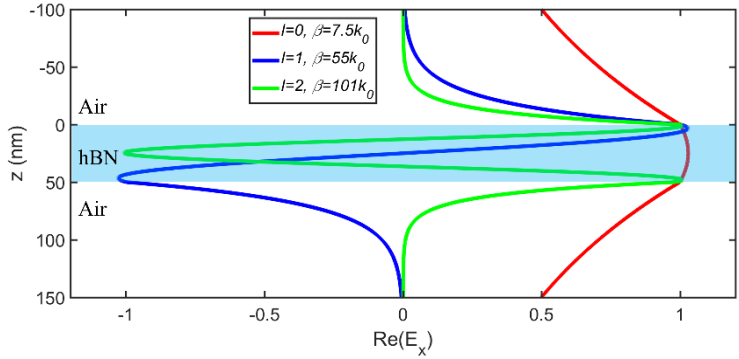

Fig. 5 (a) Normalized amplitude distribution of electric field along the $x$-axis at angular frequency of $2.8 \times 10^{14}$ rad/s.

Finally, we discuss the influence of the distance on the coupling between HPPs of two hBN sheets. Figs. 6(a) and 6(b) shows the photon tunneling probability when the distance is 10 nm and 50 nm, respectively. The thickness of each hBN sheet is 5 nm. Compared Fig. 6(a) with Fig. 2(b), one can see that 0-order HPP contributes more to the photon tunneling probability, while the 1-order HPP can also contributes to the photon tunneling probability. The reason is that the 1-order HPP in hBN sheet can only propagate limited distance in the air, and the HPPs in the hBN sheets can couple only when the distance between them is smaller than the propagating distance.

However, when the distance is 50 nm, as shown in Fig. 6(b), only small-wavevector HPPs can contribute to the photon tunneling probability. The reason is that the distance between two hBN sheets is much larger than the propagation length of the high-wavevector HPPs.

Here only the type II hyperbolic band is considered, and similar phenomenon can occur in the type I hyperbolic band, and the mechanism are the same. Besides, from Fig. 5 one can see that the strongest electric field is not at the interface of air/hBN, indicating that the hyperbolic phonon polaritons are internal polaritons [30].

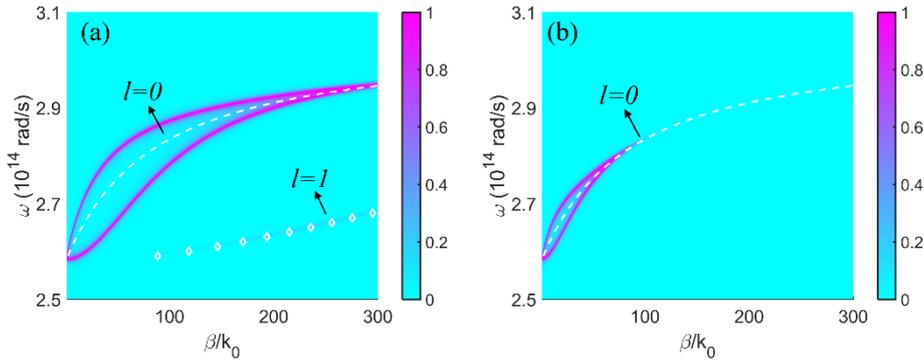

Fig. 6 Photon tunneling probability as a function of the wavevector and the angular frequency when the thickness is 5 nm: (a) $d$=10 nm; (b) $d$=50 nm. The white dashed line corresponds to the 0-order HPPs, while the white discrete diamond corresponds to the 1-order HPPs.

## 3. Conclusions

We have investigated the HPPs excited in ultrathin hBN sheets and the coupling HPPs between two hBN sheets. The dispersion relation and distribution of electric field are calculated to confirm the excitation of HPPs. When the distance between the two hBN sheets is smaller than the propagation length of the HPPs in the air, the excited HPPs in the hBN sheets can strongly couple and enhance the photon tunneling probability.

The splitting of the HPPs is similar to that of the surface waves, and such phenomenon is well explained in this work. We believe that this work will deepen our understanding of the HPPs in ultrathin hyperbolic materials. In addition, the knowledge about the HPPs will help us understand the near-field radiative heat transfer between hyperbolic materials.

**Acknowledgements**

This work was supported by the National Natural Science Foundation of China (No. 52106099), the Natural Science Foundation of Shandong Province (No. ZR2020LLZ004).